# The Key to Organisational and Construction Excellence: A Study of Total Quality Management

[*1]Mubarak Reme IBRAHIM, [2]Douglas Umbugala MUHAMMAD, [3]Bala MUHAMMAD, [4]John Alaezi OKWE & [5]Jummai AGIDANI

[12345]Department of Building, Faculty of Environmental Sciences, Baze University Abuja, Nigeria

* Corresponding Author: mubarak.ibrahim@bazeuniversity.edu.ng

## Abstract

The construction industry has shifted towards adopting Total Quality Management (TQM) techniques in order to maintain a competitive advantage. Despite prior research exploring the relationship between TQM and organisational performance, the need for a tailored approach to performance objectives is an overwhelming ambition. This study aims to examine the impact of TQM practices on organisational outcomes by surveying the knowledge of contractors in the construction industry. The study identifies eight key TQM practices and assesses their impact on established performance indicators using a questionnaire administered to 275 contracting organisations in Abuja, Nigeria. The results reveal a significant relationship between TQM practices such as top executive commitment, education and teaching, process control, and continuous progress, and the success of contractor organisations. The study also uses mapping to demonstrate how these TQM practices can be leveraged to enhance performance outcomes. These findings emphasis' the need for flexible and adaptive implementation of TQM practices, rather than a one-size-fits-all approach, in order to achieve organisational performance improvement. Additionally, the study highlights the importance of ongoing commitment from top management in promoting the significance of TQM throughout the organisation.

## 1.0 Introduction

The construction industry, despite being a major contributor to the economy of a nations, has faced persistent issues that hinder its growth and development(Alaloul et al., 2021; Boadu et al., 2020). Fragmentation, instability, low productivity, poor quality, and lack of standards have been long-standing problems that have attracted various criticisms, including poor performance in terms of profitability, operational performance, customer satisfaction(Al-Sabek, 2015; Sadikoglu & Olcay, 2014), investments in innovative research and development, and productivity improvement (Proverbs et al., 2000). The drive for construction contractors is further threatened by the changing business landscape, with rapidly changing technology (Robinson et al., 2004), democratisation of information, globalisation, and government pressure playing a significant role.

As educated clients demand superior quality products and services, organisations in the construction industry have resorted to adopting Total Quality Management (TQM) practices in order to take performance to the optimal (Arditi & Gunaydin, 1997; Love et al., 2000; Pheng & Teo, 2004; Wong, 1999, Umbugala 2016). While the implementation of TQM in other industries, including manufacturing and services, has been successful in generating enhanced products and services, increased customer satisfaction, reduced costs, improved financial performance, increased competitiveness, and raised productivity (Fotopoulos & Psomas, 2010; Tawfik Mady, 2009; Terziovski & Samson, 1999), its impact on the construction industry remains largely unknown. This is due to the unique characteristics that distinguish the construction industry from other industries, which calls to question the applicability of the results from other industries to the construction sector.

In light of the above, this research seeks to examine the extent to which TQM practices and organisational performance are correlated and how TQM practices influence organisational performance indicators within the context of contractors in the construction industry. A review of existing literature on the topic highlights the lack of empirical evidence on the impact of TQM on the construction industry, making this research even more relevant with the drive towards the circular economy. By exploring the correlation and influence of TQM on performance indicators, this study aims to provide valuable insights into the effectiveness of TQM practices in the construction industry and to contribute to a better understanding of how TQM can drive the continuous improvement of organisational performance and excellence. Overall, this research is of significant importance as it aims to fill a crucial gap in the existing body of knowledge and provide a comprehensive examination of the impact of TQM practices on organisational performance within the context of the construction industry.

## 2.0 Literature Review

### 2.1 Effect of Construction Industry Performance on Economic Development

The construction sector plays a vital role in the development and growth of both developing and developed nations (Alaloul et al., 2021). It is a driving force behind the process of industrial and metropolitan growth, as seen since the beginning of the Industrial Revolution (Rostow, 2003). The sector involves a wide range of players, including professionals, tradesmen, manufacturers, trade unions, investors, local authorities, specialists, contractors, and many others, who work together to achieve successful results. The coordination of these entities requires a high level of expertise and management skills to ensure effectiveness and the efficient execution of projects.

Construction projects are significant investments, both in terms of finances and time, and therefore, require a comprehensive understanding of the construction process. This includes knowledge of the latest technologies and materials, regulations, construction practices, and market trends. The construction sector also contributes to the socio-economic development of a country by generating employment opportunities, stimulating economic activity, and creating new infrastructure that supports economic growth (Oladinrin et al., 2012).

### 2.2 Total Quality Management (TQM)

The Total Quality Management (TQM) movement has been shaped and influenced by a number of industry gurus, including Crosby, Juran, Taguchi, and Deming. Each of these individuals brought a unique perspective to the concept of quality, which has helped to evolve TQM into a comprehensive and integrated approach to management. Crosby's approach focused on the zero-defect program, where the goal was to eliminate defects in the production process and ensure that the final output met the customer's specifications. On the other hand, Juran's approach focused on identifying the root cause of quality issues, while Deming stressed the importance of 14 quality objectives in TQM implementation. Taguchi's contribution to TQM was his emphasis on the idea that any deviation from the customer's specifications results in failure and that the organisation must strive to understand and satisfy those specifications.

Quality management, in the context of TQM, goes beyond just the output or service itself. Instead, it focuses on the processes used to achieve TQM, using management tools and techniques to ensure consistent quality in all products and services (Ahire & O'shaughnessy, 1998; Prajogo & Sohal, 2006). This is achieved through the integration of quality development, quality maintenance, and quality improvement across all levels of an organisation.TQM is an effective strategy that leverages the collective efforts of management, employees, suppliers, and customers to achieve customer satisfaction and exceed expectations (Feigenbaum, 1991). It's a philosophy of continuous improvement that drives organisations to optimize their products, services, and processes. The key to success with TQM is the involvement of all stakeholders, who work together towards a common goal of providing quality products and services.

## 2.3 Concepts of Total Quality Management TQM

The implementation of Total Quality Management (TQM) is challenged by the lack of a universally accepted definition of quality (Eng Eng & Yusof, 2003). Over the years, several quality gurus such as Deming, Crosby, Feigenbaum, Oakland, and Juran have each proposed their own interpretations of quality. For example, Deming defines quality as satisfying customer expectations, while Crosby views it as conformance to specifications. (Feigenbaum, 1991) considers quality as the composite of marketing, engineering, and maintenance efforts to meet customer expectations. On the other hand, (Juran & De Feo, 2010) defines quality as fitness for purpose and (Oakland, 2012) sees it as meeting customer needs.

TQM, as defined by Turner & Muller (2004), is a holistic approach that aims to harness the efforts of all employees to eliminate errors, minimize costs, and meet client needs (Yusuf et al., 2007). The concept of TQM was developed by Dr. W. Edward Deming in the 1940s, following the conclusion of World War II. In recent years, the need for TQM has increased due to intense global competition, increased consumer awareness of quality, rapid technological advancements, and the need for organisations to attain world-class status. To meet these demands and improve their quality standards, many organisations are implementing TQM programs and quality initiatives to gain a competitive advantage and improve business performance(Talib et al., 2011). TQM is a top-down approach led by top management with a strong commitment to quality.

## 2.4 Benefits of TQM

The advancements in Total Quality Management (TQM) have resulted in various methods, perspectives, and contexts, leading to different categorizations of benefits. While (Oakland, 2014) focused on the operational improvements brought by TQM, (Goldratt & Cox, 1984) looked at the external factors such as competitiveness, customer satisfaction, and financial outcomes. Rust et al., 1995 ) summarized the main benefit of TQM as "customer satisfaction through continuous improvement in every organisational process involved in delivering products and services. Santos & Escanciano, 2002) added that this satisfaction involves both internal and external customers.

Antony et al., 2002 ) conducted a study on the critical success factors for TQM in the Hong Kong industry and identified seven benefits using factor analysis, including increased employee involvement, enhanced communication, improved productivity, improved quality, increased customer satisfaction, reduced cost of poor quality, and improved competitive advantage. The classification of TQM benefits can vary, from a general "improved customer satisfaction" to a more detailed classification into employee benefits, productivity improvements, quality enhancements, customer satisfaction, cost savings, and competitiveness gains.

## 2.5 The Relationships Between TQM Practices and Performance

This section is summarising the results of previous studies that have investigated the relationship between Total Quality Management (TQM) practices and various organisational outcomes. The authors noted that many studies have found positive associations between TQM and outcomes such as productivity and manufacturing performance, quality performance, employee satisfaction/performance, innovative performance, customer satisfaction/results, competitive edge, market share, financial performance, and aggregate firm performance (Brah et al., 2000; Demirbag et al., 2006; Fotopoulos & Psomas, 2010; Joiner, 2007; Kumar et al., 2009; Prajogo & Brown, 2004; Prajogo & Sohal, 2006; Rahman & Bullock, 2005; Talib et al., 2013; Terziovski & Samson, 1999; York & Miree, 2004; Zehir et al., 2012).

However, the authors also acknowledged that some studies have reported negative results, which suggests that the relationship between TQM and organisational outcomes is not always straightforward and may vary depending on the context and the specific TQM practices being used (Brah et al., 2002; Ittner & Larcker, 1997; McCabe & Wilkinson, 1998; Mohrman et al., 1996; Nair, 2006; Rungtusanatham et al., 1998; Sadikoglu & Olcay, 2014; Yeung & Chan, 1998). This highlights the need for further research to gain a more comprehensive understanding of the impact of TQM on organisational outcomes.

### 2.5.1 Top Management Commitment

Organisations led by top management fully committed to TQM practices and principles are more likely to experience positive outcomes in various aspects of the organisational performance. This includes improved operational performance, inventory management performance, employee performance, innovation performance, social responsibility and customer results, financial performance, and overall firm performance(Criado & Calvo-Mora, 2009; Goetsch & Davis, 2014).

Studies have shown that top management commitment to TQM practices leads to a more collaborative and participatory organisational culture, where employees are encouraged to get involved in decision-making and to take ownership of their work(Ahire & O'shaughnessy, 1998; Al-Swidi & Mahmood, 2012; Jabeen et al., 2014; Joiner, 2007; Macinati, 2008; Parast & Adams, 2012; Phan et al., 2011; Powell, 1995; Prajogo & Sohal, 2006; Samson & Terziovski, 1999). This creates a more engaged and motivated workforce, which, in turn, contributes to better organisational performance and excellence. Top management commitment also supports the development of a comprehensive TQM system, which includes effective communication channels, employee development programs, and efficient information management. All of these elements work together to create a more effective and efficient organisation, where quality is integrated into every aspect of the business.

**H1: Top management commitment positively influences organisational performance.**

### 2.5.2 Customer Focus

A strong focus on customer needs and expectations within Total Quality Management (TQM) practices leads to significant improvements in organisational performance across various key performance indicators, including operational performance, inventory management performance, employee performance, innovation performance, customer satisfaction, sales, and overall firm performance. This is supported by numerous studies(Ahire & O'shaughnessy, 1998; Al-Swidi & Mahmood, 2012; Granerud & Rocha, 2011; Jabeen et al., 2014; Joiner, 2007; Phan et al., 2011; Samson & Terziovski, 1999; Zehir et al., 2012). By focusing on meeting the needs and expectations of customers through TQM practices, organisations are able to improve the quality and reliability of their products and services, increase efficiency and productivity, and ultimately drive customer satisfaction, sales, and overall firm performance.

**H2: Customer focus positively influences organisational performance.**

### 2.5.3 Employees/People Management

Employee involvement in quality management efforts is crucial to the success of TQM initiatives. Employees should understand their role in the organisational goals and strategies to enhance performance. Empowering employees to engage in quality efforts and fostering a sense of ownership leads to better results, as employees take a vested interest in improving products/services and processes. Research has shown that effective people management positively impacts organisational performance, including operational performance, supply management performance, social responsibility and customer results, financial performance, and overall firm performance(Ahire & O'shaughnessy, 1998; Jabeen et al., 2014; Macinati, 2008; Parast & Adams, 2012; Sadikoglu & Olcay, 2014; Samson & Terziovski, 1999).

**H3: Employee/People Management positively influences organisational performance.**

## 2.5.4 Supplier Quality Management
The relationship between Total Quality Management (TQM) and supply chain management is crucial. TQM helps to reduce and streamline the supplier base to manage supplier relationships and develop strategic alliances with suppliers. It is important to involve suppliers early in the product development process to benefit from their expertise and capabilities. Supplier quality management is vital as it provides reliable and high-quality inputs that result in high-quality products and/or services(Ahire & O'shaughnessy, 1998; Jabeen et al., 2014; Phan et al., 2011; Powell, 1995). Previous studies have shown that effective supplier quality management positively impacts operational performance, inventory management, innovative performance, and overall firm performance. Thus, the hypothesis is that supplier quality management positively influences organisational performance.

**H4: Supplier quality management positively influences organisational performance.**
## 2.5.5 Education and Training
TQM organisations should provide training and education to their employees to improve their skills in their job responsibilities. Effective training leads to better quality management, improved employee performance, increased customer satisfaction, and overall improved firm performance. Studies have shown that training has a positive impact on operational performance, inventory management, employee performance, innovation, market and financial performance, and overall firm performance(Fuentes et al., 2006; Phan et al., 2011; Sadikoglu & Olcay, 2014). However, some studies have reported negative or insignificant results. The researchers propose the hypothesis that training positively impacts organisational performance.

**H5: Education/Training positively influences organisational performance.**

## 2.5.6 Process Management
Effective knowledge and process management are key to a successful Total Quality Management (TQM) strategy. Process management focuses on activities, not just results, and is proactive in reducing variations in the process and improving product quality. Effective knowledge management ensures that employees have access to reliable and consistent data and information to perform their jobs efficiently. Monitoring data on quality helps improve the turnover rate of materials and supply, identify and fix errors in processes, and minimize negative environmental impacts. Previous studies have found that knowledge, process management, and statistical control/feedback have a positive impact on operational performance, supply management, innovation, customer results, competitiveness, financial performance, and overall organisational performance(Forza & Filippini, 1998; Macinati, 2008; Parast & Adams, 2012; Phan et al., 2011; Sadikoglu & Olcay, 2014; Zehir et al., 2012).

**H6: Process management positively influences organisational performance.**

## 2.5.7 Continuous Improvement
The concept of continuous improvement aims to improve business processes, meet and exceed customer requirements, and comply with regulatory standards. In the construction industry, this involves tracking the cost of quality processes, using quality tools, continuously reviewing and improving safety and workplace environment, encouraging project quality improvement discussions, and benchmarking processes. Studies have shown that continuous improvement has a positive impact on operational performance, inventory management performance, customer satisfaction, and market performance(Al-Swidi & Mahmood, 2012; Ittner & Larcker, 1997; Macinati, 2008; Phan et al., 2011). However, it may not have a statistically significant relationship with perceived performance in the computer industry.

**H7: Continuous Improvement is positively related to organisational performance.**

### 2.5.8 Information and Analysis

Continuous Improvement focuses on reducing process variability and improving output performance by continuously reviewing and enhancing business processes. This approach involves tracking the cost of quality processes, utilizing quality tools, practicing continual review of construction safety and processes, and benchmarking processes for improvement. The results of continuous improvement efforts can positively impact operational performance, inventory management, customer satisfaction, market performance, and overall company performance.TQM philosophy emphasizes analyzing information on customer needs and operational problems to drive quality excellence. Many TQM techniques help organisations process information for improvement(Forza & Filippini, 1998; Jabeen et al., 2014; Macinati, 2008; Phan et al., 2011; Zehir et al., 2012). However, information and analysis can also inhibit organisational performance in certain cases. Empirical research has shown that information and analysis can improve customer results, supply management performance, operational performance, innovation performance, financial performance, and overall company performance.

**H8: Information and analysis positively influences organisational performance.**

### 3.0 Research Methodology

This study used deductive research design, which involved developing hypotheses and designing a research strategy to test the formulated hypotheses. The quantitative approach was employed to collect data using questionnaires to identify Total Quality Management (TQM) practices, organisational performance indicators, assess the influence of the identified TQM practices on the performance of Construction Contractors, and map out applicable TQM practices to improve the performance of Construction Contractors. Abuja, one of the construction hubs in Nigeria where most construction companies execute their projects, was chosen as the study area. The study employed convenience sampling, and the population size was 965. The questionnaire was designed to collect data from companies that were tax compliant. The collected data were grouped into four parts to collect information on the demographics of construction firms, TQM practices, performance level of construction firms based on the influence of TQM practices, and influence of practices on the established performance indicators. A total of 275 questionnaires were self-administered on a one-to-one basis, and to improve the response rate, measures such as shorter questionnaire pages and promising a summary report of the result of the study were incorporated within the survey. The targeted respondents were the executive, senior, and middle management staff. Descriptive ~~was used~~ statistics was used to describe the characteristics of the data and Pearson correlation coefficient ~~used~~ in order to determine the relationship between these variables.

### 4.0 Results and Discussions

### 4.1 Demographic of the respondents

The results of the survey showed that 52.1% of the respondents held top management positions and 46.0% held middle management positions, highlighting the level of knowledge about their organisations' performance. Additionally, 25.8% of the respondents held Diplomas and over 60% held first-degree certificates, with the remaining 14.1% having Master's degrees. With regards to experience, 15.3% of the respondents had less than five years, 39.3% had between six to ten years, and 45.4% had over 11 years of experience in the construction industry.

The importance of this distribution lies in capturing the various levels of experience in the field. It is noteworthy that over 45% of the respondents had more than 11 years of experience in the construction industry. However, the results also showed that only 22.1% of the respondents' organisations had quality certification, indicating that the level of awareness about quality certification in the Nigerian construction industry is low, which requires improvement to address the quality issues in the industry.

Table 4.1: Demographic of the respondents

| Respondents 'Distribution | Frequency | Percentage(%) |
| --- | --- | --- |
| **Managerial position:** | | |
| Top management | 85 | 51.0 |
| Middle management | 78 | 49.0 |
| | | |
| **Education level:** | | |
| Diploma | 42 | 25.8 |
| Bachelors degree | 98 | 60.1 |
| Masters degree | 23 | 14.1 |
| | | |
| **Services carried out:** | | |
| Building construction | 84 | 52.0 |
| Civil engineering | 42 | 25.2 |
| Others | 36 | 22.1 |
| | | |
| **Working experience:** | | |
| 1-5 years | 25 | 15.3 |
| 6-10 years | 64 | 39.3 |
| 11 years and above | 74 | 45.3 |
| | | |
| **Certification to Quality system:** | | |
| No | 127 | 77.9 |
| Yes | 36 | 22.1 |

## 4.2 Reliability of the research Instrument

Cronbach's alpha measures the internal consistency and reliability of a set of items that make up a single scale. It can also be calculated for any subset of items within the scale. According to (Gliem & Gliem, 2003), a newly developed measure can be considered valid if its Cronbach's alpha is above 0.60, with the ideal threshold being 0.70. A value of 0.80 or higher is considered a strong indicator of reliability. The results of the Cronbach's alpha calculation for the TQM practices and performance variable are summarized in Table 4.2. With a Cronbach's alpha of 0.890, the TQM practices show a high level of reliability. The performance variable has a Cronbach's alpha of 0.878, which is within the acceptable range, indicating that the individual constructs used to measure performance are reliable. These results confirm that the instrument used is highly reliable.

Table 4.2: Reliability Analysis

| Variable | No. Of Items | Alpha Reliability Coefficient |
|---|---|---|
| TQM practices (Independent variable) | 8 | .890 |
| Performance Measures (Dependent variable) | 7 | .878 |

Source: Field Survey (2022)

## 4.3 Descriptive Statistics for the Study Variables

Table 4.3 presents the results of the descriptive statistics for the TQM dimensions and the characteristics of the TQM dimensions and organisational performance variables studied. The results show that the mean of the TQM dimensions ranges from 3.31 to 3.82. The mean of the top management commitment is the highest (3.81), indicating that the top management plays a crucial role in leading the organisation's quality efforts. On the other hand, the mean for employee/people management is the lowest (3.31), which suggests that employees do not place much importance on performance efforts. This could be a result of the nature of the industry where the workforce is sometimes nomadic, reducing the contribution to organisational performance. The scale used in this study ranges from 1 to 4.

For organisational performance, the variables range from 2.95 to 3.86, with a scale of 1 representing 50% and 5 representing over 95%. This indicates that contractor organisations place a strong emphasis on performance variables to achieve their objectives. All of the means of the variables in this study are above the midpoint, which means that most respondents have similar opinions about each of the variables. Furthermore, the standard deviation (SD) is less than one, indicating that the variations in respondent opinions are small. In summary, mean and SD were used to determine the extent of spread of the data.

Table 4.3: Results of descriptive statistics of overall TQM practices and Organisational performance

| Variables | Mean | Std.Deviation |
|---|---|---|
| Top Management Commitment | 3.82 | .512 |
| Customer Focus | 3.72 | .503 |
| Employee/People Management | 3.31 | .641 |
| Supplier Quality Management | 3.73 | .510 |
| Education/Training | 3.70 | .620 |
| Process Management | 3.67 | .629 |
| Continuous Improvement | 3.71 | .552 |
| Information and Analysis | 3.71 | .577 |
| | | |
| Customer Satisfaction | 2.97 | .780 |
| Employee Performance | 3.66 | .611 |
| Profit Improvement | 3.84 | .541 |
| Process improvement | 2.98 | .757 |
| Market/Financial Performance | 3.86 | .516 |
| Operational Performance | 3.37 | .625 |
| Innovation Improvement | 2.95 | .776 |
| | | |
| ValidN | 163 | |

## 4.4 Correlation between TQM practices and Organisational performance variables

The study investigated the relationship between Total Quality Management (TQM) practices and organisational performance. The TQM practices include Top Management Commitment (TMC), Customer Focus (CF), Employee/People Management (E/PM), Supplier Quality Management (SQM), Education/Training (E/T), Process Management (PM), Continuous Improvement (CI), and Information and Analysis (IA). Meanwhile, the organisational performance dimensions consist of Customer Satisfaction (CS), Employee Performance (EP), Profit Improvement (PI), Process Improvement (PIm), Market/Financial Performance (M/FP), Operational Performance (OP), and Innovation Improvement (II). The results are presented in Table 4.2.

The correlation analysis showed that there is a positive relationship between the TQM practices and the organisational performance dimensions. This suggests that implementing TQM practices can lead to improvement in organisational performance in various areas, such as customer satisfaction, employee performance, financial performance, and operational efficiency. The results indicate that a focus on TQM practices can have a positive impact on organisations and contribute to their overall success.

Table 4.4: Correlation between TQM Practices variables and Organisational Performance variables

| Variables | 1 | 2 | 3 | 4 | 5 | 6 | 7 | 8 | 9 | 10 | 11 |
|---|---|---|---|---|---|---|---|---|---|---|---|

| | 1 | 2 | 3 | 4 | 5 | 6 | 7 | 8 | 9 | 10 | 11 | 12 | 13 | 14 |
|---|---|---|---|---|---|---|---|---|---|---|---|---|---|---|
| 1 TopManagementCommitment | - | | | | | | | | | | | | | |
| 2 CustomerFocus | .248** | - | | | | | | | | | | | | |
| 3 Employee/PeopleManagement | .209** | .277** | - | | | | | | | | | | | |
| 4 SupplierQualityManagement | .236** | .047 | .277** | - | | | | | | | | | | |
| 5 Education/Training | .259** | .297** | .201** | .204** | - | | | | | | | | | |
| 6 ProcessManagement | .244** | .087 | .269** | .067 | .253** | - | | | | | | | | |
| 7 ContinuousImprovement | .232** | .277** | .247** | .269** | .195* | .247** | - | | | | | | | |
| 8 InformationandAnalysis | .218** | .231** | .207** | .214** | .195* | .062 | .248** | - | | | | | | |
| 9 Customer Satisfaction | .156* | .257** | .186* | .068 | .151* | .160* | .191* | .269** | - | | | | | |
| 10 EmployeePerformance | .141* | .041 | .185* | .185* | .188* | .196* | .221** | .211** | .061 | - | | | | |
| 11 ProfitImprovement | .121* | .204** | .157* | | .185* | .180* | .220** | .267** | .069 | .288** | .204** | | | |
| 12 ProcessImprovement | .169* | .200* | .189* | .189* | .262** | .231** | .200* | .282** | .548** | .620** | .396** | | | |
| 13 Market/Financial Performance | .140* | .200* | .097 | .157* | .105 | .047 | .267** | .185* | .285** | .306** | .090 | | | |
| 14 OperationalPerformance | .169* | .101 | .234** | .127 | .203** | .235** | .285** | .044 | .417** | .105 | | | | |
| 15 InnovationImprovement | .183* | .154* | .246** | .185* | .211* | .200* | .062 | .204** | .667** | .581** | .069 | | | |

*. Correlation is significant at the 0.05 level (2-tailed), **.
Correlation is significant at the 0.01 level (2-tailed). Source: Field Survey(2022)

### 4.4.1 Top Management Commitment (TMC) and Organisational Performance dimensions

TMC was found to be positively and significantly correlated with CS (r = .156*, p< 0.05); EP (r = .141*, p< 0.05); PI (r = .121*, p< 0.05); PIm (r = .169*, p< 0.05); M/FP (r = .140*, p< 0.05); OP (r = .169*, p< 0.05); II (r = .183*, p< 0.05). Furthermore, the results showed the overall TMC has a positive relationship with organisational performance, with II, PIm, OP, and CS having the highest correlation values. In general, the finding demonstrated TMC is correlated with the organisational performance of construction firms, therefore hypothesis (H1) is supported (H1: Top management commitment positively influences the organisational performance) developed in this study.

### 4.4.2 Customer Focus (CF) and Organisational Performance variables
The relationship between CF and CS, EP, PI, PIm, M/FP, OP II was found to be positively and significantly correlated with CS (r = .257**, p< 0.05); EP (r = .041, p< 0.05); PI (r = .204**, p< 0.05); PIm (r = .200*, p< 0.05); M/FP (r = .200*, p< 0.05); OP (r = .101, p< 0.05); II (r = .154, p< 0.05).The correlation result suggests that two out of the seven organisational performance variables are significantly correlated with CF, with two more also significantly correlated and the rest; EP, OP, and II are a have weak positive correlation and not significant with CF. Furthermore, the results revealed that CF overall has a positive relationship with organisational performance, with the highest correlation values in CS, PI, PIm, and M/FP. This indicates that these four variables are strongly influenced by CF. In general, the finding indicates that the CF is correlated with organisational performance in construction organisations. Thus, hypothesis (H2) is supported, relationship (H2: Customer focus positively influences organisational performance).

### 4.4.3 Employee/People Management (E/PM) and Organisational Performance variables

The correlation results for E/PM indicated positive and significant relationship with CS (r = .186*, p< 0.05); EP (r = .185*, p< 0.05); PI (r = .157*, p< 0.05); PIm (r =.189*, p< 0.05); M/FP (r = .097, p< 0.05); OP (r = .234**, p< 0.05); II (r = .246**, p< 0.05).The correlation value suggests that two out of the seven organisational performance variables have higher correlation values with E/PM, with four having relatively higher correlation values with E/PM, while PIm has no significant relationship with E/PM. Furthermore, the results showed that E/PM overall has a positive relationship with organisational performance, with the highest correlation values being OP and II. This implied that the two variables are strongly correlated with E/PM. Overall, the finding indicated E/PM is correlated with organisational performance in construction organisations, and consequently, the hypothesis (H3) is supported (H3: Employee/People management positively influences organisational performance)

### 4.4.4 Supplier Quality Management (SQM) and Organisational Performance variables

The correlation values for SQM showed to be positively and significantly related to CS (r = .068, p< 0.05); EP (r = .185*, p< 0.05); PI (r = .185*, p< 0.05); PIm (r =.189*, p< 0.05); M/FP (r = .157*, p< 0.05); OP (r = .127, p< 0.05); II (r = .185*, p< 0.05).The results of the correlation analysis showed that organisational performance dimensions are positively correlated with SQM, but CS and OP are not significantly correlated with SQM. Furthermore, the result indicate overall SQM has a positive relationship with organisational performance, with the highest values in EP, PI, PIm, and II. This means all variables are significantly related to SQM. In general, the finding indicates that the SQM is related to organisational performance in construction organisations, thereby, the hypothesis (H4) is supported (H4: Supplier quality management positively influences organisational performance).

### 4.4.5 Education/Training (E/T) and Organisational Performance variables

The correlation values for E/T showed were positively and significantly correlated with CS (r = .151, p< 0.05); EP (r = .188*, p< 0.05); PI (r = .180*, p< 0.05); PIm (r = .262**, p< 0.05); M/FP (r = .105, p< 0.05); OP (r = .203**, p< 0.05); II (r = .211**, p< 0.05). The finding revealed the overall E/T has a positive relationship with organisational performance, with the highest correlation values in PIm, OP, and II. These variables are correlated with E/T, and therefore, the supported hypothesis (H5) (H5: Education/Training positively influences the organisational performance) developed in this study.

### 4.4.6 Process Management (PM) and Organisational Performance variables

The correlation value for PM was found to be positively and significantly correlated with CS (r = .160*, p< 0.05); EP (r = .196*, p< 0.05); PI (r = .220**, p< 0.05); PIm (r =.231**, p< 0.05); M/FP (r = .047, p< 0.05); OP (r = .235**, p< 0.05); II (r = .200*, p< 0.05). The finding indicated organisational performance dimensions are positively correlated with PM, but market/financial performance is not significantly correlated with PM. Furthermore, the result demonstrated overall PM has a positive relationship with organisational performance, with the highest correlation values being PI, PIm, OP, and II. This means the variables are strongly correlated with PM. In general, the results showed that the PM influences organisational performance in construction organisations, and therefore, the hypothesis (H6) is supported (H6: Process management positively influences organisational performance)

### 4.4.7 Continuous Improvement (CI) and Organisational Performance variables

The correlation value for CI was found to be positively and significantly correlated with CS (r = .191*, p< 0.05); EP (r = .221**, p< 0.05); PI (r = .267**, p< 0.05); PIm (r =.200*, p< 0.05); M/FP (r = .267**, p< 0.05); OP (r = .285**, p< 0.05); II (r = .269**, p< 0.05). The results also revealed that overall CI is significantly positively correlated with organisational performance, with the highest correlation values in EP, Profit Improvement, PIm, M/FP, OP, and II. In general, the finding indicates that the CI is correlated with organisational performance in construction organisations. In line with this hypothesis, (H7) is supported (H7: Continuous Improvement is positively related to organisational performance).

### 4.4.8   Information and Analysis (IA) and Organisational Performance variables

The correlation value for IA was found to be positively and significantly correlated with CS (r = .269**, p< 0.05); EP (r = .211**, p< 0.05); PI (r = .069, p< 0.05);   PIm (r =.282**, p< 0.05); M/FP (r = .185*, p< 0.05); OP (r = .268**, p< 0.05); II (r =   .204**, p< 0.05). The findings showed that organisational performance dimensions are positively correlated with IA, but profit improvement is not significantly related to IA. Furthermore, the results showed that the overall IA has a positive relationship with organisational performance, with the highest correlation values in CS, EP, OP, PIm, and II. In general, the findings indicated that IA is correlated with organisational performance in construction organisations. Thus, hypothesis (H8) is supported (H8: Information and analysis positively influences organisational performance).

The results of the correlation indicated the extent of relationship between the two groups of variables but did not identify the most crucial variables for these relationships. To achieve this, multiple regressions were conducted between TQM practices and OP. This is to determine the importance of each independent variable and its contribution to the relationship.

### 4.5 Strategies for Improving Organisational Performance

The results of the mean ranking show the level of influence each TQM practice has on organisational performance. The practices were evaluated on a scale of 1 to 4, with 1 representing "Highly Insignificant" and 4 representing "Highly Significant." This helps to provide a clear picture of the impact each practice has on the overall performance of the organisation.The mapping of the results, as depicted in Figure 4.1, showcases the significance of each TQM practice. The practices were streamlined and their influence was analyzed, which has helped to identify the most impactful practices. This information is valuable for organisations as they can use it to prioritize their TQM efforts and allocate resources more effectively. The results of this mean ranking are a valuable tool for organisations looking to improve their performance through TQM. By focusing on the practices that have the greatest impact, organisations can achieve significant improvements in efficiency, customer satisfaction, and overall performance. The TQM practices with a ranking of 4 can be considered the cornerstone of any organisation's efforts to improve performance, while practices with lower rankings can be refined and improved over time.

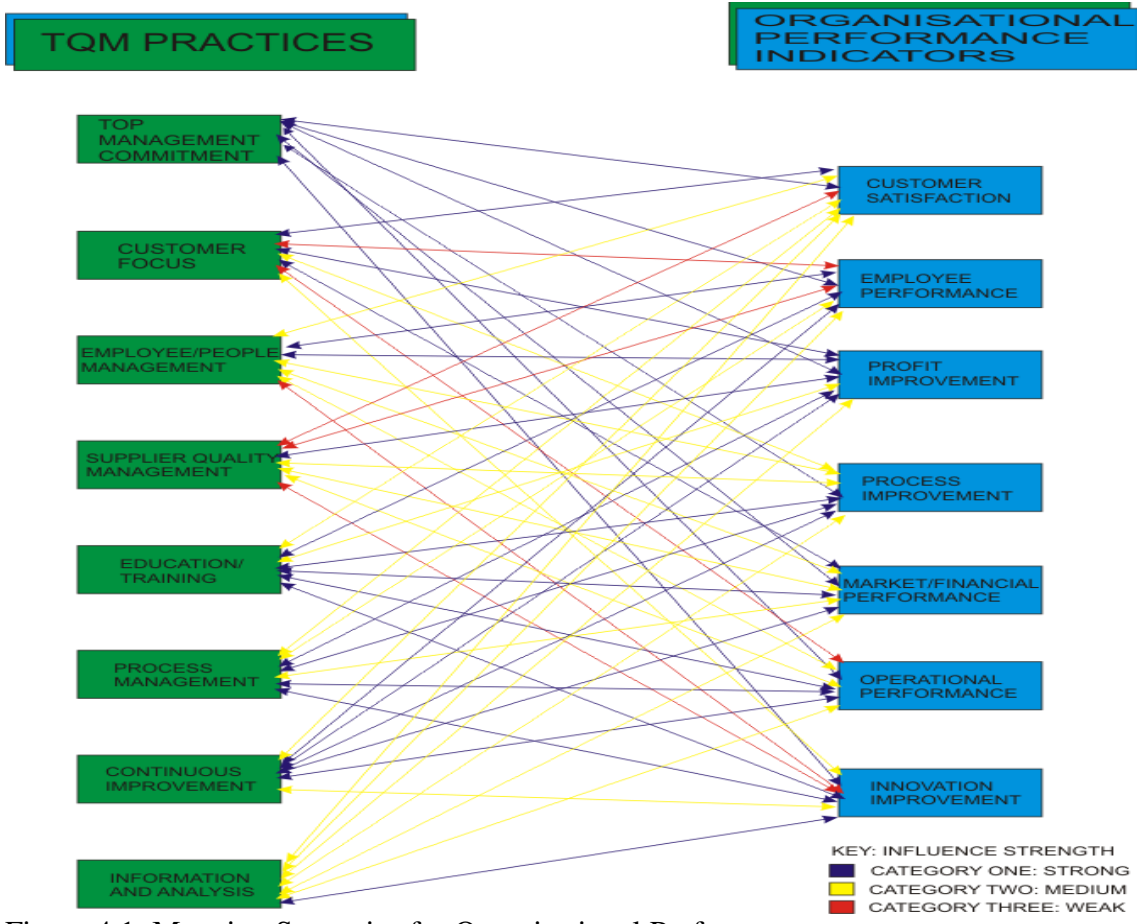

Figure 4.1: Mapping Strategies for Organisational Performance

**4.4.1   Strategy 1: Customer satisfaction performance**

i. Top management commitment commitment
ii. Customer <u>Focus</u> management
iii. Employee/people management
iv. Education/training
v. Process <u>Management</u>
vi. Continuous improvement
vii. Information and analysis
viii. Supplier quality management

**4.4.2   Strategy 2: Employee performance**

i. Top management
ii Employee/people management
iii. Education/training
iv. Continuous improvement
v. Process management
vi. Information and analysis
vii. Customer focus
viii. Supplier quality management

**4.4.3   Strategy 3: Profit improvement improvement**

i. Top management commitment commitment
ii. Customer focus
iii. Supplier quality management
iv. Process management
v. Continuous improvement management
vi. Employee/people management
vii. Education/training

**4.4.4   Strategy 4: Process**

i. Top management
ii. Process management
iii. Education/training
iv. Continuous improvement
v. Employee/people
vi. Supplier quality management
vii. Information and analysis

viii.    Information and analysis

viii.    Customer focus

**4.4.5    Strategy 5: Market Performance performance**
  i.    Top management commitment commitment
  ii.   Customer focus
  iii.  Education/training
  iv.   Continuous improvement
  v.    Employee/people management management
  vi.   Supplier quality management
  vii.  Process management
  viii. Information and analysis

**4.4.6    Strategy 6: Operational**
  i.    Top management commitment
  ii.   Education/training
  iii.  Process management
  iv.   Continuous improvement
  v.    Employee/people
  vi.   Supplier quality management
  vii.  Information and analysis
  viii. Customer focus

**4.4.7    Strategy 7: Innovation improvement**
  i.    Top management commitment
  ii.   Education/training
  iii.  Process management
  iv.   Information and analysis
  v.    Customer focus
  vi.   Employee/people management
  vii.  Continuous improvement
  viii. Supplier quality management

**5.0 Summary, Conclusion and Recommendation**
**5.1 Summary**
The study found positive and significant relationships between Top Management Commitment (TMC), Customer Focus (CF), Employee/People Management (E/PM), Supplier Quality Management (SQM), and Education/Training (E/T) with various dimensions of Organisational Performance in construction firms. The highest correlation values were found between TMC and Customer Satisfaction (CS), Process Improvement (PI), Process Implementation (PIm), and Organisational Performance (OP). CF had the highest correlation values with CS, PI, PIm, and Market/Financial Performance (M/FP). E/PM had the highest correlation values with OP and Institutionalisation (II). SQM had the highest correlation values with EP, PI, PIm, and II. E/T had the highest correlation values with PIm, OP, and II. These findings suggest that all five variables positively influence the organisational performance in construction firms.

**5.2 Conclusion**
It can be concluded that there is a significant positive relationship between Top Management Commitment (TMC), Customer Focus (CF), Employee/People Management (E/PM), Supplier Quality Management (SQM), and Education/Training (E/T) and Organisational Performance dimensions in construction firms. This indicates that a construction firm that prioritizes these factors is likely to have better performance outcomes compared to one that does not. In order to improve organisational performance in the construction industry, it is recommended that firms focus on strengthening TMC, CF, E/PM, SQM, and E/T. For example, by investing in employee training and education, a firm can improve its overall knowledge and skills base, which can lead to better customer service and improved supplier quality management that will meet sustainability and dynamic of times. Additionally, by committing to customer focus, a firm can ensure that its services are aligned with customer needs and expectations, which can lead to improved customer satisfaction and loyalty. By focusing on these key areas, construction firms can improve their overall performance and competitiveness.

## 5.0 Recommendation

The study could be replicated in different countries or regions to see if the findings hold up in different cultural set up and if there are any differences in the impact of the variables on organisational performance. Longitudinal study: A longitudinal study could be conducted to determine the dynamic relationship between the variables and organisational performance over time. This could provide a deeper understanding of the factors that contribute to sustained performance improvement. Future research could examine additional variables that may influence the relationship between the variables and organisational performance, such as technology adoption, market competition, and government regulations.